\documentclass[utf8]{frontiersFPHY} % for Science, Engineering and Humanities and Social Sciences articles

\setcitestyle{square} % for Physics and Applied Mathematics and Statistics articles
\usepackage{url,hyperref,lineno,microtype,subcaption}
\usepackage[onehalfspacing]{setspace}
\usepackage{bm}
\usepackage{amsthm}
\usepackage{amsmath}
\usepackage{amssymb}
\usepackage{graphicx}

\usepackage{bm}% bold math

\def\keyFont{\fontsize{8}{11}\helveticabold}
\def\firstAuthorLast{Franovi{\'c} {et~al.}} %use et al only if is more than 1 author
\def\Authors{Igor Franovi{\'c}\,$^{1,*}$, Sebastian Eydam\,$^{2,*}$, Serhiy Yanchuk\,$^{3,4}$ and Rico Berner\,$^{3,5,6,*}$}
\def\Address{$^{1}$Scientific Computing Laboratory, Center for the Study of Complex Systems, Institute of Physics Belgrade, University of Belgrade, Pregrevica 118, 11080 Belgrade, Serbia\\
$^{2}$Neural Circuits and Computations Unit, RIKEN Center for Brain Science, 2-1 Hirosawa, 351-0198 Wako, Japan\\
$^{3}$Institut f\"ur Mathematik, Technische Universit\"at Berlin, Strasse des 17. Juni 136, 10623 Berlin, Germany\\
$^{4}${Potsdam Institute for Climate Impact Research, 14473 Potsdam, Germany}\\
$^{5}$Institut f\"ur Theoretische Physik, Technische Universit\"at Berlin, Hardenbergstr. 36, 10623 Berlin, Germany\\
$^{6}$Institut f\"ur Physik, Humboldt-Universit\"at zu Berlin, Newtonstraße 15, 12489 Berlin, Germany}
\def\corrAuthor{}
\def\corrEmail{}

\begin{document}

\onecolumn
\firstpage{1}
\title[Resource adaptation induced collective activity bursting]{Collective activity bursting in networks of excitable systems adaptively coupled to a pool of resources}

%\rem{IF}{I would suggest "bursting" instead of "bursts", because bursting indicates a persistent activity}
%{Collective activity bursts induced by resource adaptation}

\author[\firstAuthorLast]{\Authors} %This field will be automatically populated
\address{} %This field will be automatically populated
\correspondance{Rico Berner\\ rico.berner@physik.hu-berlin.de\\
	Igor Franovi{\'c}\\ franovic@ipb.ac.rs \\
	Sebastian Eydam\\ richard.eydam@riken.jp} %This field will be automatically populated

%\extraAuth{}% If there are more than 1 corresponding author, comment this line and uncomment the next one.
\extraAuth{}

\maketitle
\begin{abstract}
We study the collective dynamics in a network of excitable units (neurons) adaptively interacting with a pool of resources. The resource pool is influenced by the average activity of the network, whereas the feedback from the resources to the network is comprised of components acting homogeneously or inhomogeneously on individual units of the network. Moreover, the resource pool dynamics is assumed to be slow and has an oscillatory degree of freedom. We show that the feedback loop between the network and the resources can give rise to collective activity bursting in the network. To explain the mechanisms behind this emergent phenomenon, we combine the Ott-Antonsen reduction for the collective dynamics of the network and singular perturbation theory to obtain a reduced system describing the interaction between the network mean field and the resources. 
\tiny
\keyFont{ \section{Keywords:} local and collective excitability, heterogeneous neural populations, metabolic resources, collective bursting, adaptive coupling, switching dynamics, multiscale dynamics, multistability}
%\rem{IF}{Note the changes to the keywords} %All article types: you may provide up to 8 keywords; at least 5 are mandatory.
\end{abstract}

\section{Introduction}
% adaptive networks
% \cite{Berner2021c,Gkogkas2021,Gerstner2014,zhou2006dynamical,Popovych2013,PlotnikovLehnertFradkovEtAl2016,ZHA15a,KAS17,Diaz-Pier2016,Berner2021,gross2008adaptive,Clopath2010,Masuda2007,Butz2013,Berner2019,Berner2021e,Sorrentino2008a,Feketa2019a,Abbott2000,Lucken2016,Rohr2019,Nowke2018,Gross2006}
% \cite{Berner2021f} \cite{Sorrentino2010}

Complex dynamical networks are indispensable for modeling many processes in nature, technology, and social sciences \cite{BOC06a,STR01a,ARE08,YAN21}. In realistic situations, collective dynamics in such networks is affected by the constraints on available resources from the environment~\cite{KRO21,ROB14}, resulting in complex dynamical phenomena, especially if the systems are self-organized to operate close to criticality \cite{LEV07}. Often, additional resource dynamics gives rise to adaptive mechanisms such as frequency adaptation~\cite{KRO21,TAY10,FUH02,HA17}, delay adaptation~\cite{FIE15a,PAR20}, or various forms of homeostatic plasticity in neuronal systems~\cite{ZIE18}. 

Dynamical networks with resource constraints have been in the focus of recent studies  \cite{KRO21,ROB14,TAY10,SON20a,VIR16c,NIC17}. In particular, in \cite{SON20a} it has been investigated how phase synchronization between the mutually uncoupled system elements depends on the interaction with the environment. A mini-review \cite{ROB14} has highlighted the importance of reciprocal coupling between neuronal activity and metabolic resources in self-organizing and maintaining neuronal operation near criticality, and has also presented a general slow-fast formulation for the case where resources change slowly relative to neural activity. In \cite{VIR16c}, a discrete two-layer model has been proposed to describe a mechanism by which metabolic resources are distributed to neurons via glial cells. 
An example of frequency adaptation in Kuramoto model was provided in \cite{TAY10}, reproducing certain phenomena that are not qualitatively accounted for the classical Kuramoto model, such as long waiting times before reaching synchronization. In \cite{NIC17}, neuronal dynamics and nutrient transport were assumed to be bidirectionally coupled, such that the allocation of the transport process at one layer depends on the degree of synchronization in the other and vice versa. 
In \cite{KRO21}, a system of coupled Kuramoto oscillators that consume or produce resources depending on their oscillation frequency was considered.  
 %\cite{Park2020,LueckenRosinWorlitzerEtAl2017} 
Inspired by the mechanisms for the interaction of a neuronal network with a population of glial cells, the studies~\cite{FIE15a,PAR20,LUE17} introduced models of networks with adaptive time-delays.

Of particular interest are adaptive networks in which connectivity changes are related to intrinsic nodal dynamics \cite{BER21c,GRO08a}. For example, these types of networks can model synaptic neuronal plasticity \cite{MEI09a,MAR11b}, chemical \cite{KUE19a,JAI01}, epidemic \cite{GRO06b}, biological, and social systems \cite{HOR20}. A paradigmatic example of adaptively coupled phase oscillators gained considerable interest recently \cite{BER19,BER20,GUT11,KAS17,BER19a,FEK20,BER21b}. This type of phase oscillator models seems to be useful for predicting and describing phenomena in more realistic and detailed models~\cite{POP15,LUE16,ROE19a} as well as for the understanding of collective phenomena such as multicluster states~\cite{BER19,BER19a} or recurrent synchronization~\cite{THI22}.

In the present paper, we consider coupled excitable systems \cite{LIN04,IZH07}, characterized by a linearly stable rest state susceptible to finite-amplitude perturbations. Excitable systems act as nonlinear threshold-like elements, such that applying a sufficiently small perturbation gives rise to a small-amplitude linear response, while a perturbation exceeding a certain threshold may trigger a large-amplitude nonlinear response. A classical example for the excitability feature are neurons \cite{IZH07,ERM86} which respond to a supra-threshold stimulation by emitting a spike. Apart from neuronal systems, excitability is important for other living cells \cite{SCI21}, lasers \cite{TER20,YAN19a}, chemical reactions \cite{CHI06}, machine learning \cite{CEN19}, and many other fields. A variety of phenomena, including resonances, oscillations, patterns and waves, are caused by the interplay of excitability and noise~\cite{PIK97,ZHE18,BAC18a,BAC18b,FRA20,POT08,NEI99a,FRA15,FRA18a,BAC20} or time-delay~\cite{BRA09,KLI16b}.

As a prototype of excitable local dynamics, we consider active rotators, paradigmatic for type I excitability \cite{LIN04,FRA20,OSI07,DOL17,KLI21a,SHI86,PAR96b}. Active rotators have been used to study interacting excitable systems with noise \cite{LIN04}, synchronization in the presence of noise
\cite{DOL17,KLI21a,SHI86,PAR96b}, the interplay of noise and an adaptive feedback \cite{FRA20}, effects of an adaptive network structure~\cite{THA21}, co-effects of noise, coupling, and adaptive feedback~\cite{SON20a,BAC18b} or delayed feedback~\cite{YAN19a} and the impact of higher-order Fourier modes~\cite{RON21}, to name but a few.

An important ingredient of our model is the multiscale structure of the dynamics, whereby the processes at the pool of resources are assumed to occur much slower than the dynamics of excitable units at the nodes. Utilizing this feature, we apply the methods of singular perturbation theory~\cite{KUE15,DES12} to first study the fast dynamics (layer dynamics) for fixed resource levels with the Ott-Antonsen approach, and then reduce the problem to the slow dynamics of resources. 

Our main result is to demonstrate how the adaptive interaction between a network of excitable systems with a pool of resources gives rise to collective activity bursting. Such emergent dynamics is characterized by alternating episodes of stationary and oscillating behavior of the macroscopic order parameter. We describe the mechanisms behind the activity bursting and indicate parameter regions where this phenomenon can be reliably observed. So far, collective bursting phenomena have been considered to emerge due to time-varying neuronal inputs~\cite{STO02}, the interplay of external input and homeostatic plasticity~\cite{ZIE18}, or synaptic short-term plasticity~\cite{GAS20}. In these studies, possible implications for healthy and diseased brain states have been drawn. Moreover, the important role of bursting phenomena for the understanding of brain-organ interactions have been highlighted in the perspectives article~\cite{IVA21}. Our study complements recent research on emergent bursting dynamics in brain and organ systems by providing a simple and analytically tractable model generating collective activity bursting.

Our paper is organized as follows. In Section \ref{sec:model} we lay out the model of a heterogeneous population of excitable units adaptively coupled to a pool of resources, while in Section \ref{sec:recurrentBurst} we introduce the main phenomenon of collective activity bursting. Sections \ref{sec:layerDyn} and \ref{sec:resdyn} concern the analysis of the system's multiscale dynamics within the framework of singular perturbation theory, first elaborating on the layer problem and then using the reduced problem to explain the mechanism of collective bursting and the origin of multistability in the full system. Section \ref{sec:switch} proposes two different approaches to induce switches between the coexisting collective regimes, whereas Section \ref{sec:conclusions} provides our concluding remarks and outlook.

%\cite{ZHA21a} - not appropriate, RB: ok
%--------------------------------------------------
% Model
%--------------------------------------------------
\begin{figure}
	\centering
	\includegraphics{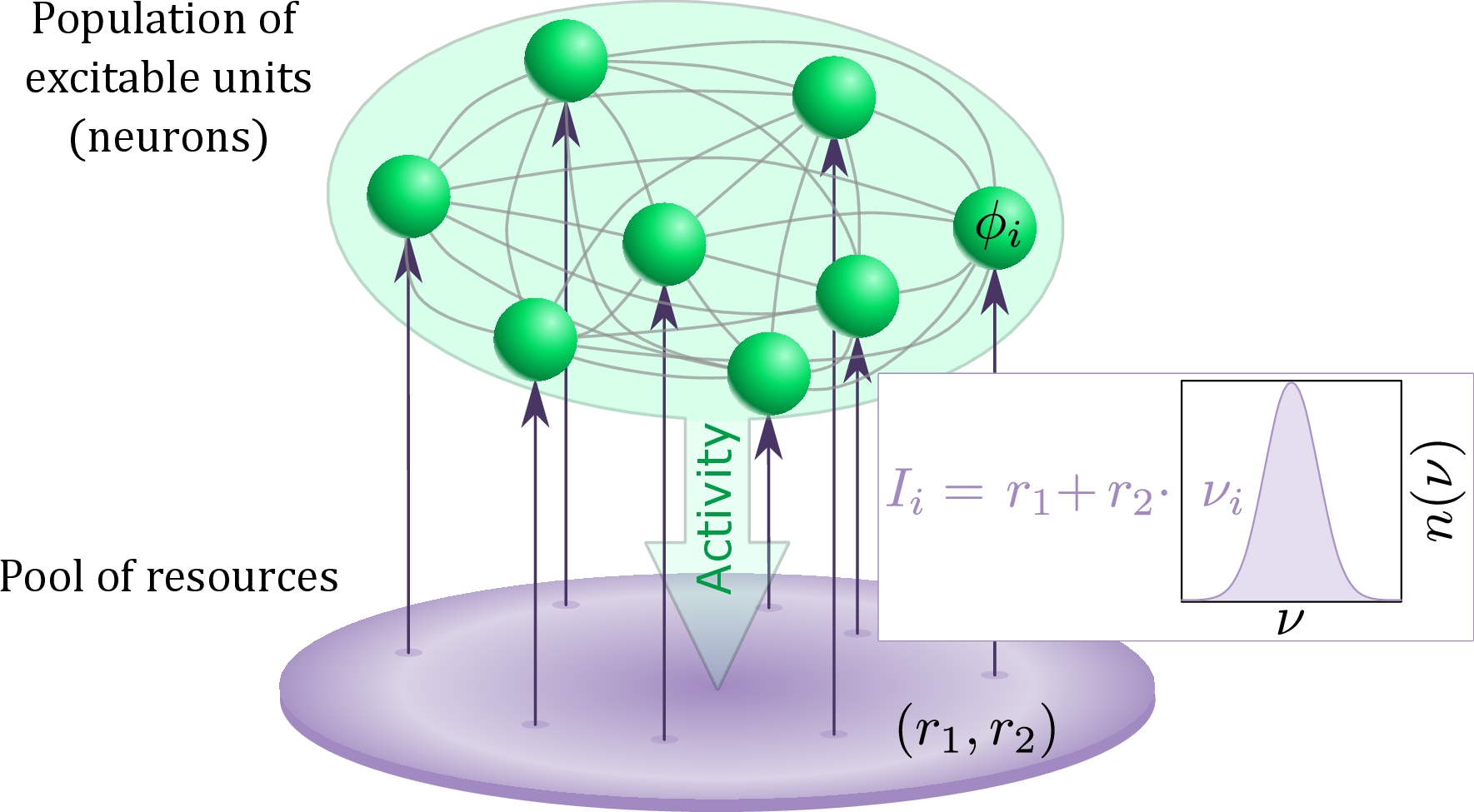}
	\caption{Schematic for the two-layer model consisting of a heterogeneous population of excitable units (green) and interacting pool of resources described by an adaptive Stuart-Landau oscillator (purple). The heterogeneity $\nu_i$ of the excitable units are randomly drawn from a distribution $n(\nu)$.}\label{fig:model}
\end{figure}

\section{Model}\label{sec:model}
%\the\textwidth
We consider a system of $N$ coupled active rotators \cite{STR94a} with a Kuramoto-type coupling given by
\begin{align}\label{eq:activeRotModel}
    \dot{\phi}_k &= I_k(\bm{r})-\sin\phi_k+\frac{\sigma}{N}\sum_{j=1}^N\sin(\phi_j-\phi_k),
\end{align}
where $\phi_k\in [0,2\pi), k=1,...,N$ are the local phase variables, and $\sigma$ is the coupling strength. While providing a simplified description of local dynamics, active rotators manifest the excitability feature crucial to neuronal activity~\cite{IZH07,STR94a} and are similar to the model of theta neurons~\cite{LAI14,LUK13}, paradigmatic for type I neural excitability. 
External inputs $I_k(\bm{r}(t))= r_1(t) + r_2(t)\nu_k$ received by each unit comprise of a \emph{homogeneous} component $r_1(t)$, acting identically at all the units, and a \emph{heterogeneous} component,
%of variance $r_2^2$,
where the variability is due to parameters $\nu_k$ drawn from a normalized Gaussian distribution $\nu_k\in\mathcal{N}(0,1)$. Recall that in models of coupled active rotators, terms $I_k$ are classically interpreted as local bifurcation parameters describing individual oscillation frequencies. Nevertheless, here $I_k(t)$ at each moment follow a Gaussian distribution $g(I)=\mathcal{N}(r_1,r_2^2)$, such that the local velocities of the units are modulated by coupling to $r_1$ and $r_2$. The latter modulation can be seen as describing an interaction with the \emph{resources} from the environment~\cite{KRO21,SON20a} summarized within the two-component resource variable $\bm{r}=(r_1,r_2)$. In the context of neuroscience such modulation of local velocities is reminiscent of frequency adaptation of neuronal spiking~\cite{FUH02,HA17} due to a limited amount of metabolic resources affecting e.g. neurotransmitters. 

Adaptation of spiking activity is a slow process compared to spike emission~\cite{ROB14,HA17}, which should be reflected in the dynamics of metabolic resources $\bm{r}(t)$. Here we propose a simple model of resource dynamics based on the Hopf normal form. We consider $\bm{r}$ as a complex variable, i.e, $\bm{r}=r_1 + \mathrm{i} r_2$, which satisfies the dynamical equation
\begin{align}
    \dot{\bm{r}} &= \epsilon f(\bm{r}-\bm{s},\lambda), \label{eq:rDyn}\\
    \dot{\lambda} &= -\epsilon' (\lambda - \lambda_0 - \gamma A(t)), \label{eq:lambdaDyn}
\end{align}
with activity
\begin{align}
    A(t) := \frac{1}{N} \sum_{j=1}^N \dot{\phi}_j,
    \label{eq:activity}
\end{align}
the metabolism describing function given by $f(\bm{r},\lambda)=\bm{r}(\lambda+\mathrm{i}\omega - |\bm{r}|^2)$, the frequency $\omega$ %\rem{SE}{Here $\omega$ is only the oscillation frequency of the Hopf normal form but the actual frequency should be $\varepsilon\omega$. Maybe we can instead say: the frequency parameter $\omega$} 
and the resource base level given by $\bm{s}=s_1+\mathrm{i} s_2$.
Small parameters $\epsilon\ll 1$ and $\epsilon' \ll 1$ are introduced to account for the scale separation between the fast spiking dynamics of units and the slowly adapting dynamics of the resources. Note that we consider the case $\epsilon'=\epsilon$ throughout the paper.

System \eqref{eq:rDyn}-\eqref{eq:lambdaDyn} that describes the interaction between the two resources undergoes a supercritical Hopf bifurcation at $\lambda=0$. This allows for the interpretation of the resource dynamics as \emph{inactive} if $\lambda<0$, when it possesses a stable focus at $\bm{s}$, and as \emph{active} if $\lambda>0$, when it displays a stable limit cycle. We further assume that the dynamics of metabolic resources is in turn adapted to the activity $A(t)$ of the population, see Eq.~\eqref{eq:activity}, 
% The population activity is defined by
% \begin{align}\label{eq:activity}
%     A(t)=\frac1N \sum_{j=1}^{N}\dot{\phi}_j,
% \end{align}
and the parameter $\gamma$ determines the adaptation strength. In case of no spiking activity, i.e., if $A(t)=0$ or $\gamma=0$, we assume that the resource dynamics is inactive and that the corresponding resource activity variable $\lambda$ settles to %$\lambda_0=-0.05$.
$\lambda_0$. 
To describe the coherence of the population dynamics, we use the complex order parameter $Z$ defined by
\begin{align}\label{eq:orderParam}
    Z(\bm{\phi}(t))=\frac{1}{N}\sum_{j=1}^N e^{\mathrm{i}\phi_j(t)} = R(\bm{\phi}(t))e^{\mathrm{i}\Theta(\bm{\phi}(t))},
\end{align} 
where $R$ is the Kuramoto order parameter, and $\Theta$ is the mean phase \cite{BIC20}.

Summarizing, we have proposed a multiscale model of a heterogeneous population of active rotators, featuring excitability and spike frequency adaptation as two important ingredients of typical neuronal activity, coupled to a pool of resources that slowly adjusts its dynamics to the activity of the population.  Figure~\ref{fig:model} provides an illustration of our model.

%--------------------------------------------------
% Activity bursting
%--------------------------------------------------
\section{Collective activity bursting}\label{sec:recurrentBurst} 
%\rem{IF}{If we call the phenomenon "collective activity bursting", then this should be reflected in the title of this section}
\begin{figure}
    \centering
    \includegraphics{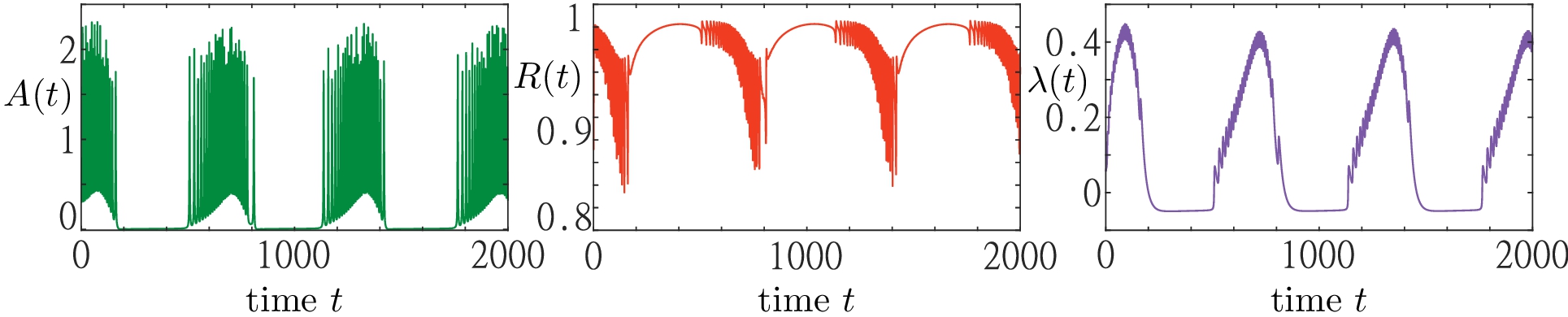}
	\caption{Collective activity bursting in system~\eqref{eq:activeRotModel}--\eqref{eq:lambdaDyn}. Three panels show the time traces of the population activity $A(t)$ (green), the order parameter $R(t)$ (red) and the resource activity variable $\lambda(t)$ (blue) from left to right, respectively. The trajectory is obtained from a random initial condition for a system of $N=5000$ active rotators and parameters: $\sigma=5$, $\epsilon=0.05$, $s_1=0.97$, $s_2=1.2$, $\omega=0.2$, $\lambda_0=-0.05$, $\gamma=0.5$.}\label{fig:phenomenon}
\end{figure}

In this section, we briefly introduce the phenomenon of collective activity bursting induced by an adaptive coupling to resources. A more detailed analysis of the phenomenon will be performed in the subsequent sections. 

In Figure~\ref{fig:phenomenon}, we show a simulation result of a system consisting of $N=5000$ active rotators adaptively coupled to a pool of resources as described by~\eqref{eq:activeRotModel}--\eqref{eq:lambdaDyn}. The emergent collective dynamics within the population is represented by the macroscopic variables $A(t)$ and $R(t)$. The dynamics within the resource pool is characterized by the activity variable $\lambda(t)$. 
We observe that the population of active rotators displays a recurrent temporal formation of bursts in the macroscopic activity $A(t)$ followed by periods of inactivity. Such episodes of macroscopic activity and inactivity correspond to episodes of a rapidly and slowly varying order parameter $R(t)$, respectively. Switching between the different regimes is equally well visible in the evolution of the resource variable $\lambda(t)$ showing the pattern of recurrent activation ($\lambda>0$) and deactivation ($\lambda<0$).

We note that this recurrent switching between macroscopic activity and inactivity is due to the adaptive feedback provided by the dynamical resources and can not be observed in a system of active rotators alone. In fact, active rotators are a paradigmatic model for excitable systems, supporting regimes of either activity $\dot{\phi}_i>0$ or inactivity $\dot{\phi}_i=0$ depending on parameters such as the input currents $I_i$, see e.g.~\cite{FRA20} for more details. The slow adaptation of the input currents caused by the resource dynamics, however, provides a mechanism to switch between the two regimes. In the following sections, we systematically describe the emergence of collective activity bursting by making use of the separation of timescales between the dynamics of the population and the resources. The slow-fast analysis within singular perturbation theory, see e.g.~\cite{KUE15,MAE15}, allows for a splitting of multiscale dynamics into a so-called layer dynamics of the fast variables and an averaged dynamics for the slow variables. 

The layer dynamics of system~\eqref{eq:activeRotModel}--\eqref{eq:lambdaDyn} consists of a network of actively coupled rotators with input currents drawn for a Gaussian distribution $\mathcal{N}(r_1,r_2^2)$. The subsequent analysis of the layer equation in Section~\ref{sec:layerDyn} provides us with a clear mapping for the regimes of population activity and inactivity. Building on this, we analyse the full system~\eqref{eq:activeRotModel}--\eqref{eq:lambdaDyn} and show that the collective activity bursting emerge close to criticality, i.e., the boundary between activity and inactivity of the layer dynamics. We also describe regimes of multistability between activity bursting and inactivity, and provide insights into perturbations that give rise to transitions between different states.

%--------------------------------------------------
% Analysis of the layer equation
%--------------------------------------------------
\section{Layer dynamics: Heterogeneous population of active rotators}\label{sec:layerDyn}

The fast subsystem, describing the evolution of the original slow-fast problem \eqref{eq:activeRotModel}--\eqref{eq:lambdaDyn} on the 
fast timescale, comprises of a heterogeneous assembly of $N$ globally coupled active rotators
\begin{align}
    \dot{\phi_k} &= r_1 + r_2\nu_k-\sin{\phi_k}+\frac{\sigma}{N}\sum\limits_{j=1}^N\sin(\phi_j-\phi_k). \label{eq:layer}
\end{align}
In the absence of adaptation of the resource variables $r_1$ and $r_2$, the local dynamics $\dot \phi_k = I_k -\sin \phi_k$ depends on external input $I_k=r_1 + r_2\nu_k$ which may be seen as an effective bifurcation parameter mediating the transition between an excitable ($I_k\lesssim 1$) and oscillatory regime ($I_k>1$) via a SNIPER (saddle-node infinite period) bifurcation at $|I_k|=1$. In the singular limit $\epsilon\rightarrow 0$, system \eqref{eq:layer} defines the layer problem, where $r_1$ and $r_2$ are treated as additional system parameters. 

According to classical singular perturbation theory~\cite{KUE15,MAE15}, the layer problem describes solutions of the multiscale system~\eqref{eq:activeRotModel}--\eqref{eq:lambdaDyn}
on a timescale much shorter than $1/\epsilon$, where the variables $r_1$ and $r_2$ do not change significantly. In particular, it can describe fast (rapidly changing) segments of the solutions. 
%The slow segments, they converge to solutions of the reduced problem~\cite{KUE15}.

\subsection{Ott-Antonsen approach for the layer dynamics}
We analyze the layer problem by determining the stability of stationary solutions of the layer dynamics and their bifurcations within the framework of Ott-Antonsen theory~\cite{OTT08,OTT09a}. We start by rewriting the layer dynamics in terms of complex order
parameter~\eqref{eq:orderParam}, which leads to
\begin{align}
    \dot{\phi}_k&=I_k-\sin{\phi_k}+\sigma \mathrm{Im}(Z(t)e^{-i\phi_k}). 
\label{eq:lay1}
\end{align}
In the thermodynamic limit $N\rightarrow\infty$, the state of the population can be described by the
probability density $h(\phi,I,t)$, which satisfies the normalization condition $\int_0^{2\pi}h(\phi,I,t)\mathrm{d}\phi=g(I)$, see e.g.~\cite{OME12b,OME13b}. The continuity equation for $h(\phi,I,t)$ then reads
\begin{align}
    \frac{\partial h}{\partial t}+\frac{\partial}{\partial \phi}(hv)=0, \label{eq:cont}
\end{align}
where the velocity is given by $v=I-\sin{\phi}+\sigma \mathrm{Im}(Z(t)e^{-i\phi})$. According to Ott-Antonsen ansatz~\cite{OTT08,OTT09a}, the long-term dynamics of \eqref{eq:cont} settles onto an invariant manifold of the form
\begin{align}
    h(\phi,I,t)&=\frac{g(I)}{2\pi}\{1+\sum\limits_{n=1}^{\infty}[\bar{z}^n(I,t)e^{\mathrm{i}n\phi}+z^n(I,t)e^{-\mathrm{i}n\phi}]\},
\label{eq:ansatz}
\end{align}
where $z(I,t)$ is the local order parameter, connected with the global complex order parameter~\eqref{eq:orderParam} via 
\begin{align}
    Z(t)=\int_{-\infty}^{\infty}g(I)z(I,t)dI. \label{eq:locgl}
\end{align}
Inserting~\eqref{eq:ansatz} into~\eqref{eq:cont}, one obtains the Ott-Antonsen equation for the layer dynamics
\begin{align}
\dot{z}&=\frac{1}{2}(1-z^2)+iIz+\frac{\sigma}{2}Z-\frac{\sigma}{2}\bar{Z}z^2, 
\label{eq:Ott}
\end{align}
where bar denotes the complex conjugate.

\subsection{Stationary solutions of the layer dynamics}
To find stationary solutions of \eqref{eq:locgl}--\eqref{eq:Ott}, we first write the local order parameter in polar form $z(I,t)=\rho(I,t)e^{i\vartheta(I,t)}$. Separating for the real and imaginary parts, equation~\eqref{eq:Ott} becomes
\begin{align}
\begin{split}
    \dot{\rho}&=\frac{1}{2}(1-\rho^2)B\cos \Phi,\\
    \rho\dot{\Phi}&=I \rho-\frac{1}{2}(1+\rho^2)B\sin \Phi, 
\end{split} \label{eq:twod}
\end{align}
where the new variables $B,\beta$ and $\Phi$ are given by
\begin{align}
\begin{split}
    B(t)e^{i\beta(t)}&=1+\sigma R(t)e^{i\Theta(t)},\\
    \Phi&= \vartheta-\beta.
\end{split} \label{eq:b}
\end{align} 
From \eqref{eq:b}, it follows that $B$ and $\beta$ are related with the macroscopic order parameter~\eqref{eq:orderParam} via
\begin{align}
\begin{split}
    B&=\sqrt{1+\sigma^2 R^2+2\sigma R \cos \Theta},\\
    \tan \beta&=\frac{\sigma  R \sin \Theta}{1+\sigma R \cos \Theta}. 
\end{split} \label{eq:brho}
\end{align}

Note that the local dynamics can be rewritten in terms of $B$ as $\dot{\phi}_k=I_k-B\sin(\phi_k-\beta)$, suggesting that $B$ may be understood as an effective excitability parameter that describes how local excitability is changed by the impact of interactions. As a consequence, the structure of stationary solutions of the Ott-Antonsen system \eqref{eq:twod} depends on the relation between $|I_k|$ and $B$, such that a population splits into two groups comprised of excitable ($|I|<B$) or oscillating units ($|I|>B$). In particular, the stationary solutions $(\rho^*,\Phi^*)$ are given by
\begin{align}
\label{eq:stst1}
\begin{split}
    (\rho^*,\Phi^*)&=\left(1,\arcsin\frac{I}{B}\right),\\
    (\rho^*,\Phi^*)&=\left(1,\pi-\arcsin\frac{I}{B}\right),
\end{split} 
\end{align}
for the excitable (inactive) group, and
\begin{align}
\label{eq:stst2}
    (\rho^*,\Phi^*)=\left(\frac{|I|-\sqrt{I^2-B^2}}{B},\frac{\pi}{2}\mathrm{sign}(I)\right)
\end{align}
for the oscillating (active) group. An explicit expression for $B$ can be obtained by invoking the self-consistency relation between the global and local order parameter
\eqref{eq:locgl}. Inserting the results for the stationary local and global 
order parameter (using \eqref{eq:locgl},  \eqref{eq:stst1}, \eqref{eq:stst2}, and the first equation from \eqref{eq:b}) and separating for the real and imaginary parts, one ultimately arrives at a self-consistency equation for $B$~\cite{KLI19,LAF10}
\begin{align}
p(B)=B^2-2\sigma p_2(B)+\frac{\sigma^2}{B^2}(p_1^2(B)+p_2^2(B))-1=0, \label{eq:Bsc}
\end{align}
where $p_1(B)$ and $p_2(B)$ are given by
\begin{align}
    p_1(B)&=r_1-\int\limits_{|I|>B} I g(I-r_1)\sqrt{1-\left(\frac{B}{I}\right)^2}\,\mathrm{d}I, \nonumber \\
    p_2(B)&=\int\limits_{|I|<B}g(I-r_1)\sqrt{B^2-I^2}\,\mathrm{d}I.
\end{align}
Having determined $B$, the stationary local and global order parameters can be obtained using the relations $R=\sqrt{p_1^2+p_2^2}/B$ and $\Theta=\arctan(p_1/(p_2-\sigma R^2))$, which follow from equations \eqref{eq:locgl} and  \eqref{eq:brho}.

\begin{figure}
	\centering
	\includegraphics[scale=0.3]{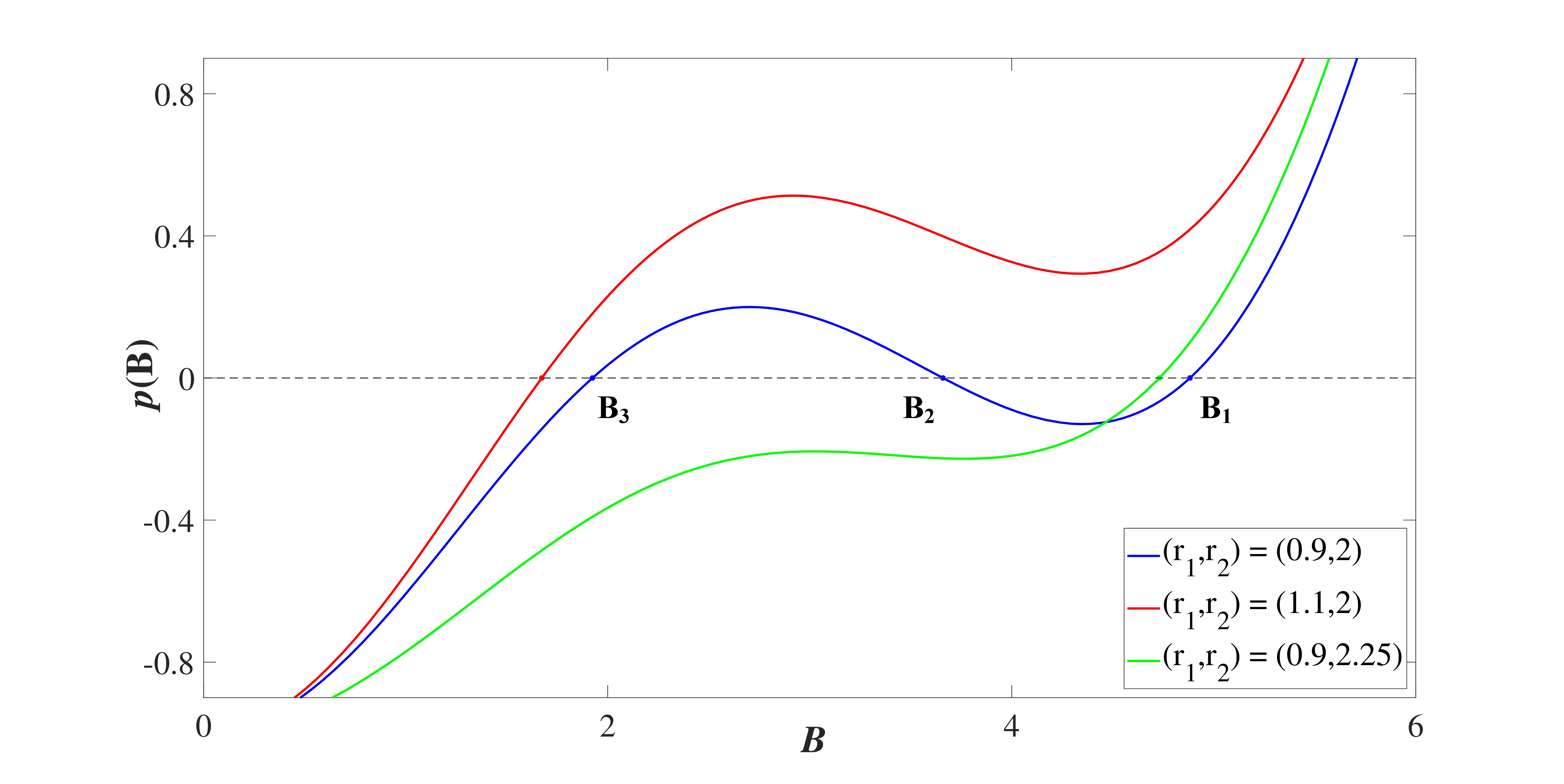}
	\caption{Changes in form and the number of roots of the function $p(B)$ given by~\eqref{eq:Bsc} under variation of $r_1$ and $r_2$ for fixed $\sigma=5$. The function $p(B)$ has three roots for $(r_1,r_2)=(0.9,2)$ (blue line; roots indicated by letters) and a single root for $(r_1,r_2)=(1.1,2)$ (red) and $(r_1,r_2)=(0.9,2.25)$ (green).
	%\rem{SY}{make color lines more thick}\rem{IF}{Done}
	}\label{fig:proots}
\end{figure}

\begin{figure}
	\centering
	\includegraphics[scale=0.35]{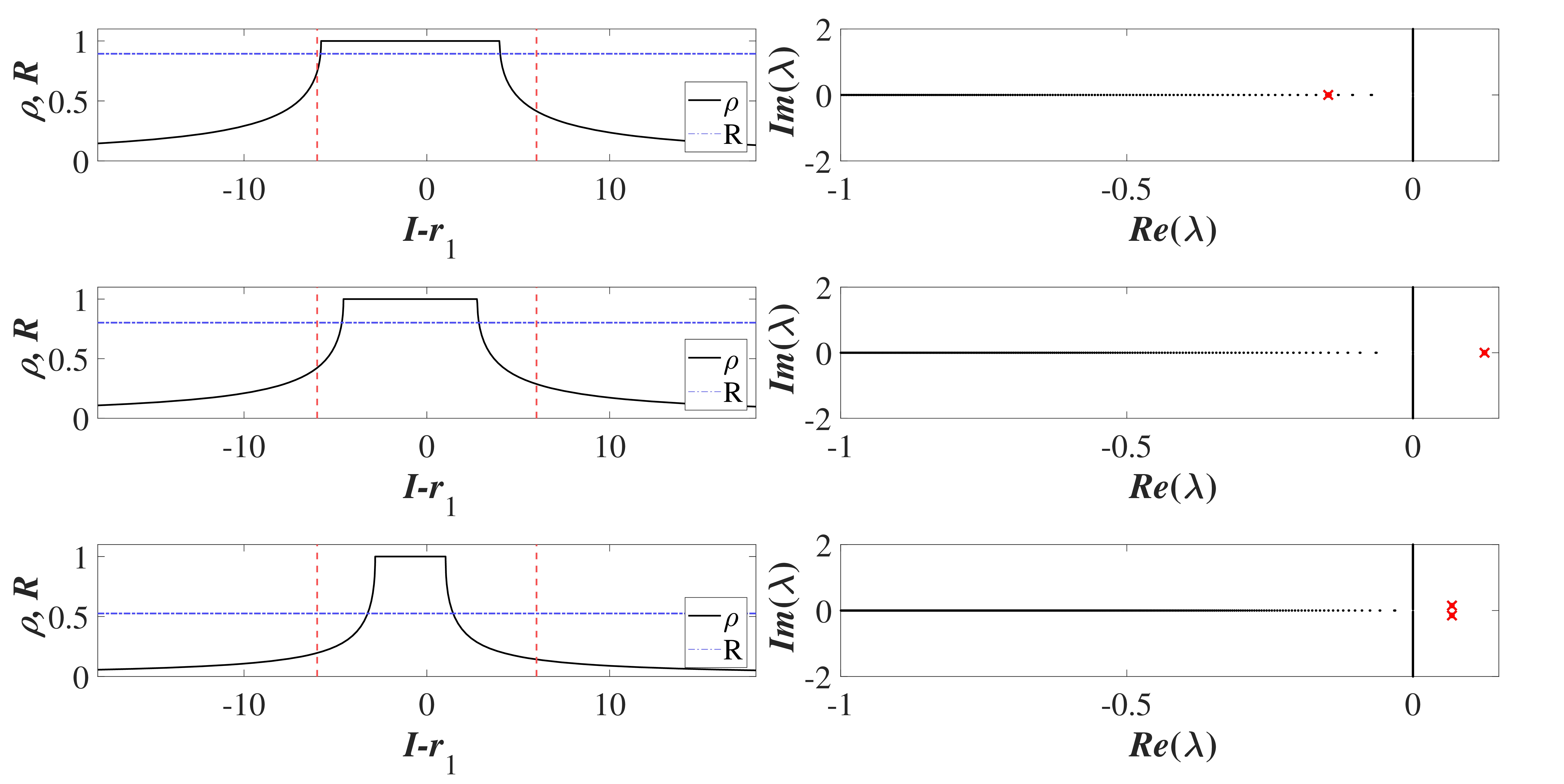}
	\caption{Local structure and spectra of stationary solutions $B_1$ (a), $B_2$ (b) and $B_3$ (c) of Ott-Antonsen equation \eqref{eq:Ott} for $\sigma=5$ and $(r_1,r_2)=(0.9,2)$. Left column shows the dependencies of the local order parameter on the input $z(I)$ (black solid lines) and the corresponding Kuramoto order parameter $R$ (blue dash-dotted lines) for the three stationary solutions. Red dashed lines indicate the interval $(r_1-3r_2,r_1+3r_2)$ relevant for the distribution of external inputs. Right column shows the continuous (black dots) and the discrete spectra (red crosses) for the stationary solutions: $B_1$ and $B_2$ are stable and unstable nodes, respectively, while $B_3$ is an unstable focus. 
	%\rem{SY}{labels for the panels (a) ,... are missing}
	%\rem{SY}{make lines more thick}\rem{IF}{Done}
	%\rem{RB}{The black dots are hard to distinguish from red ones (in particular in gray scale prints). @IF: Could you change one of them to crosses or diamonds?}
	}\label{fig:lorder}
\end{figure}

For a fixed coupling strength $\sigma$, the function $p(B)$ may have from one to three roots, depending on the mean value $r_1$ and the standard deviation $r_2$ of the distribution of intrinsic parameters $I_k$. The examples in Figure~\ref{fig:proots} illustrate how the number of solutions of~\eqref{eq:Bsc} changes between one and three for fixed $\sigma=5$, $r_1=0.97$ under increasing $r_2$. We refer to the stationary solutions by the corresponding $B$ values, which we arrange in decreasing order $B_1>B_2>B_3$. Recalling the arguments above, one sees that the larger $B$ value implies a prevalence of excitable over oscillating units within the local structure of the stationary state. This is evinced by the left column of Figure~\ref{fig:lorder} which shows the dependence of the local order parameter $z(I)$. Typically, the state $B_1$ comprises of a clear majority of excitable units, corresponding to a coherent domain $z=1$, and may thus be referred to as a \emph{homogeneous} stationary state. The two remaining stationary states $B_2$ and $B_3$ are \emph{heterogeneous} in the sense that they involve a mixture of excitable and asynchronously oscillating units. 

\subsection{Stability and bifurcation analysis of stationary solutions}
Given that Ott-Antonsen equation \eqref{eq:Ott} contains both the global order parameter and its complex conjugate, stability and bifurcation analysis of the stationary solutions~\cite{OME13b,KLI19} can be carried out by writing the local and global order parameters as $z(I,t)=x(I,t)+\mathrm{i}y(I,t), Z(t)=X(t)+\mathrm{i}Y(t)$ and separating for the real and imaginary parts. This results in the system
\begin{align}
\begin{split}
    \dot{x}&=F(x,y,X,Y)=\frac{1}{2}(1-x^2+y^2)-I y+\frac{\sigma}{2}X-\frac{\sigma}{2}X(x^2-y^2)-\sigma xyY,\\
    \dot{y}&=G(x,y,X,Y)=-xy+I x+\frac{\sigma}{2}Y-\sigma xyX+\frac{\sigma}{2}Y(x^2-y^2), 
\end{split}\label{eq:oadec}
\end{align}
which can be linearized for variations $\xi=(\delta x,\delta y)^T, \Xi=(\delta X,\delta Y)^T$ 
around the stationary solution $(x_0,y_0,X_0,Y_0)$, ultimately arriving at
\begin{align}
    \frac{\mathrm{d}\xi}{\mathrm{d}t}&=\hat{P}(I)\xi(I,t)+\hat{Q}(I)\Xi(t), \label{eq:linop}
\end{align}
where $\hat{P}$ and $\hat{Q}$ are the corresponding Jacobian matrices
$$
\hat{P}=\left(
      \begin{array}{cc}
        \frac{\partial F}{\partial x} & \frac{\partial F}{\partial y} \\
        \frac{\partial G}{\partial x} & \frac{\partial G}{\partial y} \\
      \end{array}
    \right),
    \hat{Q}=\left(
        \begin{array}{cc}
          \frac{\partial F}{\partial X} & \frac{\partial F}{\partial Y} \\
          \frac{\partial G}{\partial X} & \frac{\partial G}{\partial Y} \\
        \end{array}
      \right).
      $$
Equation \eqref{eq:linop} is augmented by the variational equation for \eqref{eq:locgl}:
\begin{equation}
\label{eq:pert-solv}
\Xi(t) = \int_{-\infty}^{\infty} g(I) \xi(I,t)\,\mathrm{d}I.    
\end{equation}
Assuming that the variations $\xi(I,t)$ and $\Xi(t)$ satisfy
$\xi(I,t)=\xi_0(I)e^{\mu t},\Xi(t)=\Xi_0 e^{\mu t}$, systems~\eqref{eq:linop} and \eqref{eq:pert-solv} transform into
\begin{align}
(\hat{P}(I)-\mu\hat{\mathbb{I}})\xi_0(I)+Q(I)\Xi_0=0,
\quad 
\Xi_0 = \int_{-\infty}^{\infty} g(I)\xi_0\,\mathrm{d}I,
\label{eq:eigenp}
\end{align}
where $\hat{\mathbb{I}}$ denotes the identity operator. From the general spectrum theory of linear
operators~\cite{OME13b,MIR07}, it follows that the Lyapunov spectrum of Eq.~\eqref{eq:eigenp} consists of a
continuous and a discrete part. Here, the continuous spectrum turns out to be always stable or 
marginally stable, %\rem{SY}{why?}\rem{IF}{This is completely system dependent, I cannot provide a qualitative argument why this should or shouldn't apply}
such that the stability of stationary solutions depends on the discrete spectrum. The latter can be determined by rewriting \eqref{eq:eigenp} in the form $\hat{C}(\mu)\Xi_0=0$, where
\begin{align}
    \hat{C}(\mu)&=\hat{\mathbb{I}}+\int_{-\infty}^{\infty} g(I)(\hat{P}(I)-\mu\hat{\mathbb{I}})^{-1}Q(I)\,\mathrm{d}I. \label{eq:spectrum}
\end{align}
The discrete spectrum is then obtained by solving the characteristic equation $\det \hat{C}(\mu)=0$ \cite{KLI19}. An example of the discrete and continuous spectra calculated for the stationary states $B_1$, $B_2$ and $B_3$ at $(r_1,r_2)=(0.9,2)$ is provided in the right column of Figure~\ref{fig:lorder}.

\subsection{Comparison between analysis and numerics}

\begin{figure}
	\centering
	\includegraphics{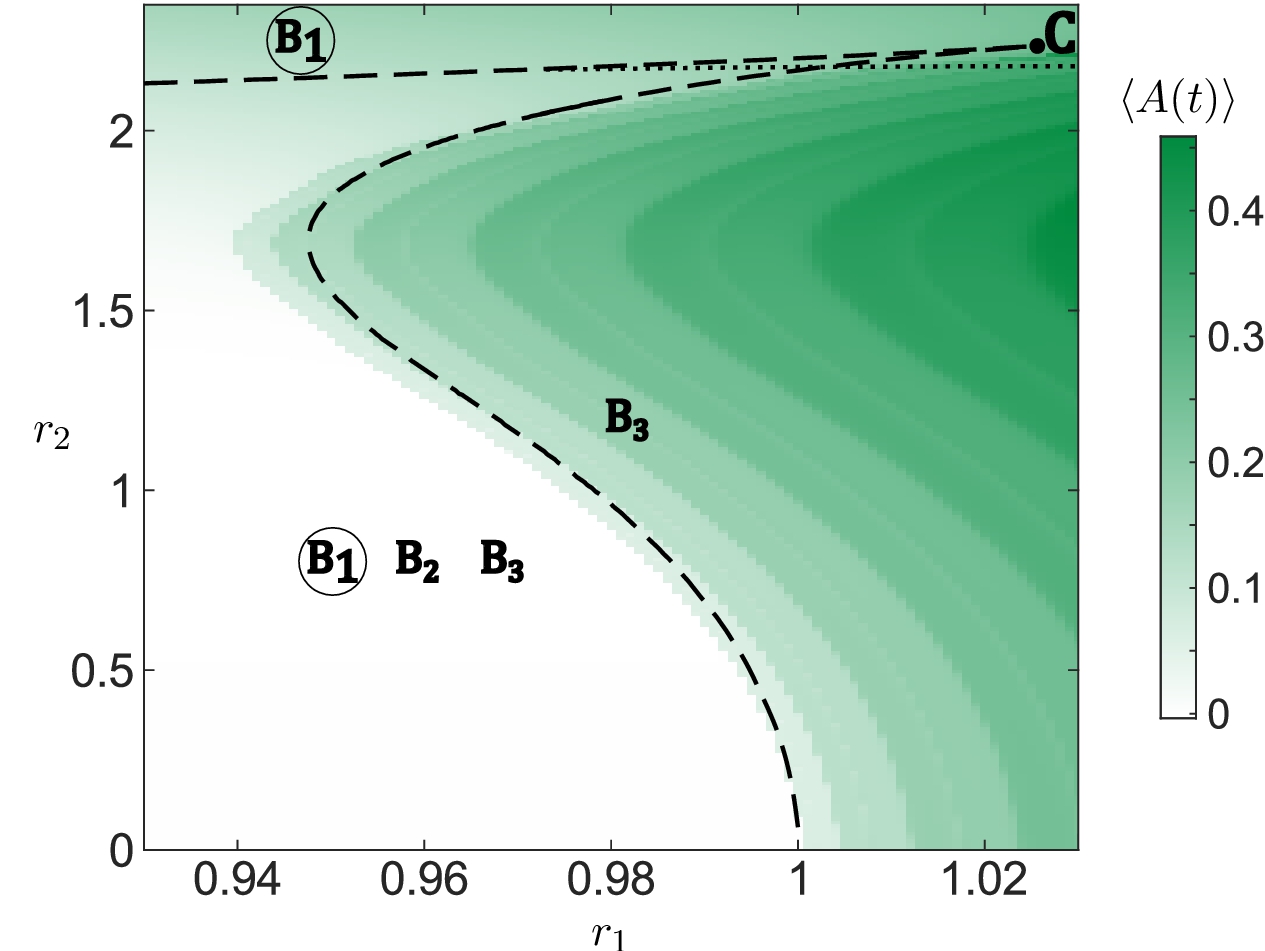}
	\caption{Bifurcation diagram for the system of active rotators~\eqref{eq:activeRotModel} in dependence on resource levels $r_1$ and $r_2$. For the simulation, we have chosen one set of random initial conditions for the phases and one set of parameters $\nu_k$ randomly drawn from a normalized Gaussian distribution $\mathcal{N}(0,1)$. Simulations comprise of $200$ time units with activity averaged over the last $100$ time units. Fold bifurcations involving stationary solutions $B_1$ and $B_2$ (lower branch) and $B_2$ and $B_3$ (upper branch) obtained from~\eqref{eq:Bsc} are shown by black dashed lines that give rise to a cusp point marked with $C$. Existence of particular solutions and their stability, derived from the discrete spectrum of~\eqref{eq:spectrum}, are indicated by their corresponding letters and a circle, respectively, whereby the circle indicates a stable solution. Along the black dotted line, stationary solution $B_3$ changes its stability in a Hopf-like bifurcation. Other parameters: $N=5000$, $\sigma=5$.}\label{fig:OAvsNumerics}
\end{figure}

The previous analysis allows for an analytic description of the existence and stability of stationary solutions in the limit of large populations ($N\to\infty$). In particular, the bifurcation diagram for the Ott-Antonsen equation of the layer dynamics~\eqref{eq:Ott} in the $(r_1,r_2)$ plane is 
organized around a co-dimension two cusp point, indicated by C in Figure~\ref{fig:OAvsNumerics} where the two branches of folds meet (black dashed lines). Both branches of folds are calculated by numerical continuation of the solutions of \eqref{eq:Bsc} using the software package BifurcationKit.jl \cite{VEL20a}. The lower branch of folds which folds over for larger $r_2$ corresponds to annihilation and reemergence of a pair of equilibria, $B_1$ and $B_2$, whereby the former (latter) is always stable (unstable). For smaller $r_2$, crossing this branch either by enhancing $r_1$ or $r_2$ gives rise to long-period collective oscillations as a stable equilibrium $B_1$ vanishes by colliding with an unstable equilibrium $B_2$. The divergence of the oscillation period when approaching the curve indicates that it corresponds to a SNIPER bifurcation of the full system. For larger $r_2$, as the branch folds over, one observes the reappearance of a stable stationary state $B_1$, emerging in an inverse fold bifurcation together with an unstable equilibrium $B_2$. The upper branch of folds involves stationary states $B_2$ and $B_3$, such that they collide and disappear above the curve, where $B_1$ remains the only stable stationary state, cf. Figure~\ref{fig:OAvsNumerics}. 

Note that apart from the fold bifurcation, the stability of $B_3$ is also affected by a Hopf-like bifurcation (black dotted line). Above the given curve, stability of $B_3$ is determined by a pair of complex conjugate eigenvalues which have the smallest negative real parts. However, crossing the curve, these eigenvalues merge with the imaginary axis and remain neutrally stable immediately below the curve%\rem{SY}{this looks non-generic, any ideas why?}\rem{IF}{It is non-generic, but also Matthias told me he already encountered a couple of times such a scenario when analyzing OA equation}
, implying that the central manifold theorem associated to Hopf bifurcation cannot immediately be applied. Still, in close vicinity below the curve, starting from an initial condition corresponding to $B_3$ results in oscillations similar to a genuine scenario of Hopf bifurcation. 

Figure~\ref{fig:OAvsNumerics} further shows a comparison of the existence and stability conditions for the collective stationary states derived from Ott-Antonsen approach for the limit $N\rightarrow \infty$ with simulations for a finite population of $N=5000$ active rotators with fixed resources $r_1$ and $r_2$. One observes that simulation results agree well with the fold bifurcation lines separating parameter regimes of low and high collective activity. The differences can be attributed to the finite size of assemblies considered in the simulations. 

\begin{figure}
	\centering
	\includegraphics{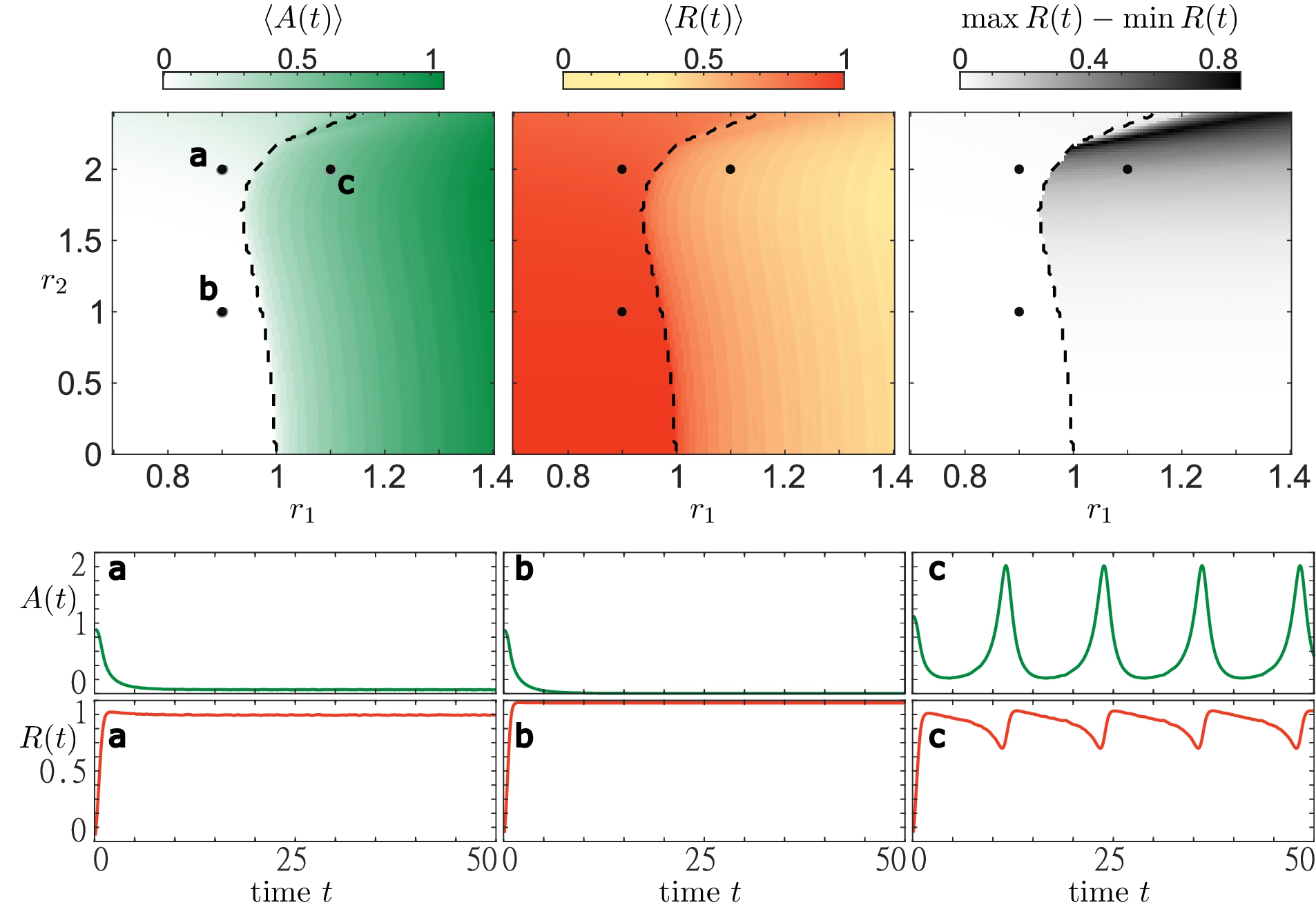}
	\caption{Bifurcation diagram for the system of active rotators~\eqref{eq:activeRotModel} in terms of resource levels $(r_1,r_2)$. All simulations are carried out the same way as in Figure~\ref{fig:OAvsNumerics}. Three diagrams at the top panel show the time-averaged values of activity $A(t)$ (white-green), order parameter $R(t)$ (yellow-red) and variations of the order parameter $\max R(t)-\min R(t)$ (white-black). The black dashed line separates regions where the mean phase $\Theta$ of the complex order parameter $Z$ features stationary or oscillating dynamics, respectively. The lower panels show time traces of activity and order parameter for three parameter pairs $(r_1,r_2)$: a - (0.9,2), b - (0.9,1), c - (1.1,2).}\label{fig:bifLayerEq}
\end{figure}
With Figure~\ref{fig:bifLayerEq} we complement the analysis of the layer equation. In particular, we show how the dynamical regimes change in a wide range of parameters $r_1$ and $r_2$,
and indicate the boundary (black dashed line) that separate parameter regions supporting stable stationary states from those admitting oscillatory states. We illustrate three different trajectories corresponding to qualitatively different collective regimes found by numerical analysis. For parameter pairs $a$ and $b$, we observe the emergence of stationary states in accordance with the bifurcation analysis shown in~Figure~\ref{fig:OAvsNumerics}. In both cases, activity $A(t)$ and the coherence measure $R(t)$ settle to a constant value. We observe that with increasing $r_2$ ($b$ to $a$) the activity level rises while the coherence level declines. For the parameter set $c$, there is no stable stationary state and we observe stable oscillations. The activity shows a regular, tonic-like spiking shape corresponding to an increase in the average activity. Meanwhile, variation of the order parameter causes its average value to decrease. In order to quantify the temporal variations of the order parameter, we also plot the difference $\max R(t)-\min R(t)$ for the considered average time interval. We observe that even though the activity level might be high, e.g. for $r_1 > 1$ close to $r_2=0$, the coherence within the population is not necessarily strongly varying. However, there are also regimes, e.g. for $r_1 > 1$ and $r_2>2$, where the order parameter varies strongly and covers almost the entire interval from $0$ to $1$.

In this section, we have illustrated the stability regions of macroscopic stationary states in a heterogeneous population of active rotators. Numerically, we have also determined the values
%oscillating states for 
of resource parameters $r_1$ and $r_2$ where no stable stationary solutions exist. Using these insights, we are now able to qualitatively describe the phenomena in systems with a slow adaptation of the resources and the resource-dependent dynamics. The next section is devoted to explaining the emerging states of collective activity bursting.

%--------------------------------------------------
% (Slow) resource dynamics
%--------------------------------------------------
\section{(Slow) resource dynamics and the emergence of multistability}\label{sec:resdyn}

The analysis of the layer dynamics in Section~\ref{sec:layerDyn} provides insight on how the system evolves for constant resources $r_1$ and $r_2$. Due to the different timescales of the population (fast dynamics) and the pool of resources (slow dynamics), we can average ~\cite{SAN07c,FRA20} the system \eqref{eq:rDyn}--\eqref{eq:lambdaDyn} as
\begin{align}
    \dot{\bm{r}} &= f(\bm{r}-\bm{s},\lambda), \label{eq:rDynAverage}\\
    \dot{\lambda} &= -\lambda + \lambda_0 + \rho \langle A\rangle, \label{eq:lambdaDynAverage}
\end{align}
where 
%we have used classical averaging theory, see e.g.~\cite{SAN07c,FRA20}, with 
$\langle A\rangle =\frac1T\int_0^T A(s)\mathrm{d}s$. Here we also assume $\epsilon'=\epsilon$ and rescale time $t_\text{new} = \varepsilon t_\text{old}$. Note that the average activity shown in Figure~\ref{fig:bifLayerEq} depends on the resource variables $\bm{r}$, since the definition~\eqref{eq:activity} implies $A=r_1 - \mathrm{Im}(Z)$. Hence, the system \eqref{eq:rDynAverage}--\eqref{eq:lambdaDynAverage} describes an effective three-dimensional coupled dynamics for the slow subsystem. A further analytical analysis of this system is beyond the scope of our study. However, we directly use the insight that the slow dynamics follows the average activity of the fast system to understand the emergence of collective activity bursting.

\begin{figure}
	\centering
	\includegraphics{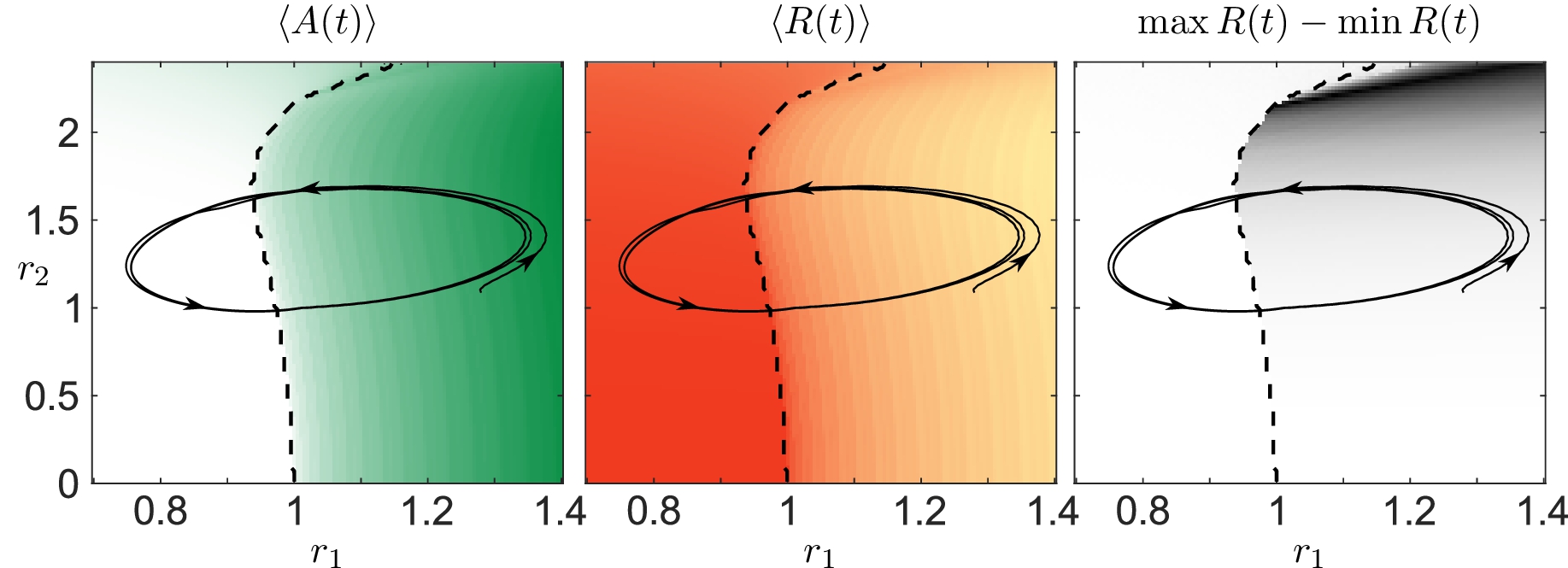}
	\caption{Bifurcation diagrams as in Figure~\ref{fig:bifLayerEq} complemented with the trajectories of the resource variables $r_1(t)$ and $r_2(t)$ for the collective activity bursting shown in Figure~\ref{fig:bifLayerEq}.}\label{fig:slowResDyn}
\end{figure}
Figure~\ref{fig:slowResDyn} shows the trajectories of the resource variables $r_1(t)$ and $r_2(t)$ for the collective bursting presented in Figure~\ref{fig:bifLayerEq} along with the averaged values of population activity and the order parameter. We clearly see that the asymptotic orbit passes through both the regimes of an active and inactive population which explains the episodes of high and low activity in Figure~\ref{fig:phenomenon}(a). Also the segments of increasing and decreasing average activity visible in Figure~\ref{fig:phenomenon}(a) can be explained by Figure~\ref{fig:slowResDyn}. Here, the average activity shows the same pattern along the trajectory ($r_1(t),r_2(t)$). The same also holds for the average values of the order parameter and even its variations if we compare Figure~\ref{fig:phenomenon}(b) with Figure~\ref{fig:slowResDyn}. Therefore, the splitting of the fast from the slow subsystem provides a very good qualitative explanation for the observed phenomenon.

To understand the emergence of the full periodic orbit shown in Figure~\ref{fig:slowResDyn}, we first note that without any population activity, i.e., $\langle A\rangle =0$ and $\lambda_0 = -0.05$, the resource dynamics possess a stable focus close to the critical line (black dashed line in Figure~\ref{fig:slowResDyn}) describing the transition from stationary to oscillatory dynamics of the mean phase $\Theta$.%\rem{IF}{What is here meant by the "critical line"? Do you refer to the fold bifurcation obtained by OA approach for the layer dynamics, because the black dashed line appears different?} \rem{RB}{The dashed line is actually one of the fold lines shown for a larger range of r1 and r2}\rem{IF,SE}{we could write: ...from stationary to oscillatory dynamics of the mean phase $\Theta$}
During the stationary phase, $r_1$ and $r_2$ tend to $s_1=0.97$ and $s_2=1.2$, respectively. As in Figure~\ref{fig:slowResDyn}, the trajectory ($r_1(t),r_2(t)$) may start in the active region, i.e., oscillatory mean phase dynamics. Due to the positive average value of activity, the variable $\lambda(t)$ characterizing the resource activity increases according
to~\eqref{eq:lambdaDynAverage} and becomes positive, see Figure~\ref{fig:phenomenon}(c). Hence, the resources become activated and ($r_1(t),r_2(t)$) follows the limit cycle solution of the resource dynamics revolving around ($s_1,s_2$). Note that the resources obey the Hopf normal form with a Hopf bifurcation at $\lambda=0$, see~\eqref{eq:rDyn}. After passing the critical line, the average activity immediately drops to $\langle A\rangle \approx 0$ which causes $\lambda$ to tend
to $\lambda_0$, see Figure~\ref{fig:phenomenon}(c). After ${\lambda}$ falls below zero, the dynamics of the resources ($r_1(t),r_2(t)$) is described by a spiral towards ($s_1,s_2$). This spiral, however, enters the active region by passing the critical line which ultimately leads to the recurrent phenomenon observed in Figure~\ref{fig:phenomenon}. As we have seen, the emergence of collective activity bursting relies on the subtle interplay between activation and deactivation of resources and the population. Furthermore, the need for the spiraling dynamics towards a stable focus explains well the necessity for the resource basis levels ($s_1,s_2$) to be close to the critical line separating the population active and inactive regimes.

With regards to the above description of the collective activity bursting, one might ask for the coexistence of a stable steady state in the system~\eqref{eq:activeRotModel}--\eqref{eq:lambdaDyn} as long as ($s_1,s_2$) lie in the inactive regime. This state might have a small basin of attraction such that the spiral towards the steady state cannot reach the active regime. 

\begin{figure}
	\centering
	\includegraphics{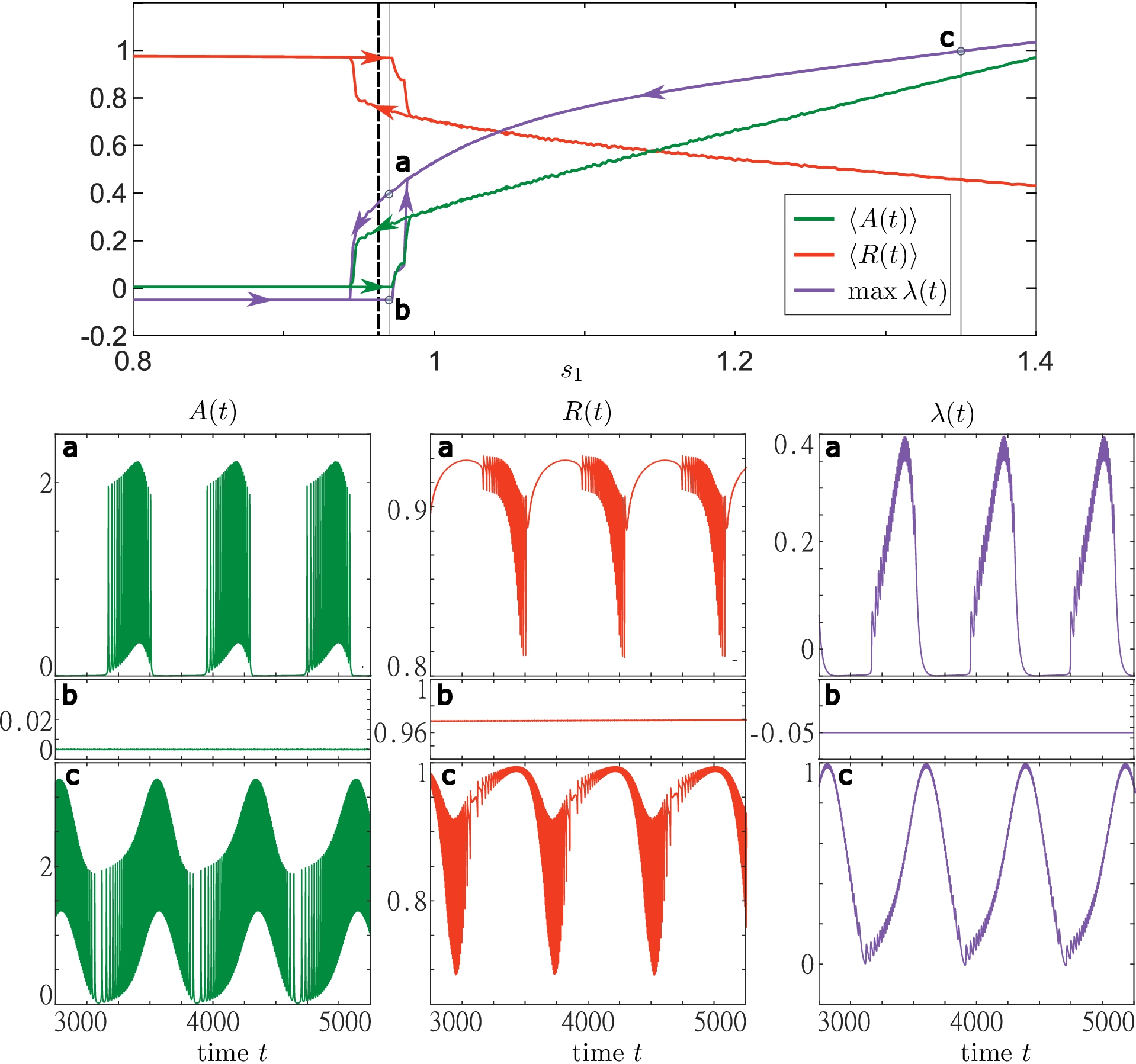}
	\caption{Bifurcation diagram with respect to the resource base level $s_1$ for a system of active rotators with adaptive resource interaction~\eqref{eq:activeRotModel}--\eqref{eq:lambdaDyn}. The top panel shows the results from two adiabatic continuations with step size $\Delta s_1=0.002$ from $s_1=0.8$ to $s_1=1.4$ (sweep up) and vice versa (sweep down). Sweeps up and down start from a stable stationary and a stable oscillatory state, respectively. For both sweeps are shown the average activity $\langle A(t)\rangle$ (green), average order parameter $\langle R(t)\rangle$ (red) and maximal resource activity $\lambda$ (blue). The results were obtained by simulating ~\eqref{eq:activeRotModel}--\eqref{eq:lambdaDyn} for $7000$ time units and taking the average over the last $5000$ time units. The branches corresponding to the two sweeps are marked by arrows. The black dashed lines indicate the value ($s_1\approx 0.963$) of the critical line shown in Figure~\ref{fig:bifLayerEq}. Three trajectories represented by the activity (green, first column), order parameter (red, middle column) and the resource activity (blue, last column) are shown in the lower panels for (a,b) $s_1=0.97$ and (c) $s_1=1.35$. The panels in (a) and (b) represent the states found along the sweeps down and up, respectively. The simulations were performed using the same values of $\nu_k$ as in Figure~\ref{fig:OAvsNumerics}. Parameters: $N=5000$, $\sigma=5$, $\epsilon=0.05$, $s_2=1.2$, $\omega=0.2$, $\gamma=0.5$.%\rem{IF}{The axis label $s_1$ in the top panel should be larger, currently it is smaller than tick labels}\rem{SE,IF}{The circle that should denote solution (c) in the top panel is missing}
	}\label{fig:sweeps}
\end{figure}
In order to get insights into the different stable states that exist in~\eqref{eq:activeRotModel}--\eqref{eq:lambdaDyn}, we use the numerical method of adiabatic continuation. To do so, we fix the base level $s_2=1.2$ and gradually vary $s_1$ from $0.8$ to $1.4$ (sweep up) and from $1.4$ to $0.8$ (sweep down). For each value of $s_1$, we run the simulation starting from the final state of the previous simulation. 

In Figure~\ref{fig:sweeps}, we show the results of both sweeps. We observe the existence of stable steady and stable oscillating states for various values of $s_1$. As expected, close to the boundary between active and inactive states of layer dynamics, we also find an interval of coexistence between collective activity bursting and stable steady states, see panels for (a) and (b) in Figure~\ref{fig:sweeps}, respectively. For larger $s_2$, only the oscillatory state can be observed, which does not enter the inactive regime above a certain $s_1$, see panel (c) in Figure~\ref{fig:sweeps}. Note that the character of the solution can be deduced from the maximal value of $\lambda(t)$ on the averaging time interval. In particular, there is a stationary state only if $\max \lambda(t) < 0$. In all other cases, there are time intervals where the trajectory of $\bm{r}$ diverges from the base level $\bm{s}$ and follows the periodic solution of~\eqref{eq:rDynAverage}.

From the arguments laid out in this section, we have seen that the mutual activation and deactivation between the neural population and the pool of resources close to criticality of layer dynamics induces a rich dynamical behavior. It is believed, particularly, that the human brain operates close to criticality~\cite{CHI10a,HAI13,YU13a,COC17,WIL19}. Therefore, it is of major importance to understand the dynamics of neural populations in this regimes including the interaction with its environment. In the next section, we propose a simple mechanism which can induce a switch between coexisting macroscopic regimes.

%--------------------------------------------------
% Switching induced by resource activation and inactivation
%--------------------------------------------------
\section{Population switching dynamics induced by resource activation and inactivation}\label{sec:switch}

In the vicinity of the transition between the population inactivity and activity, we have observed collective activity bursting induced by an adaptive dynamical pool of resources. Moreover, this phenomenon emerges in a stable coexistence with a steady state. In this section, we consider two simple perturbation approaches that can induce a switch between these two functionally different states. 

\begin{figure}
	\centering
	\includegraphics{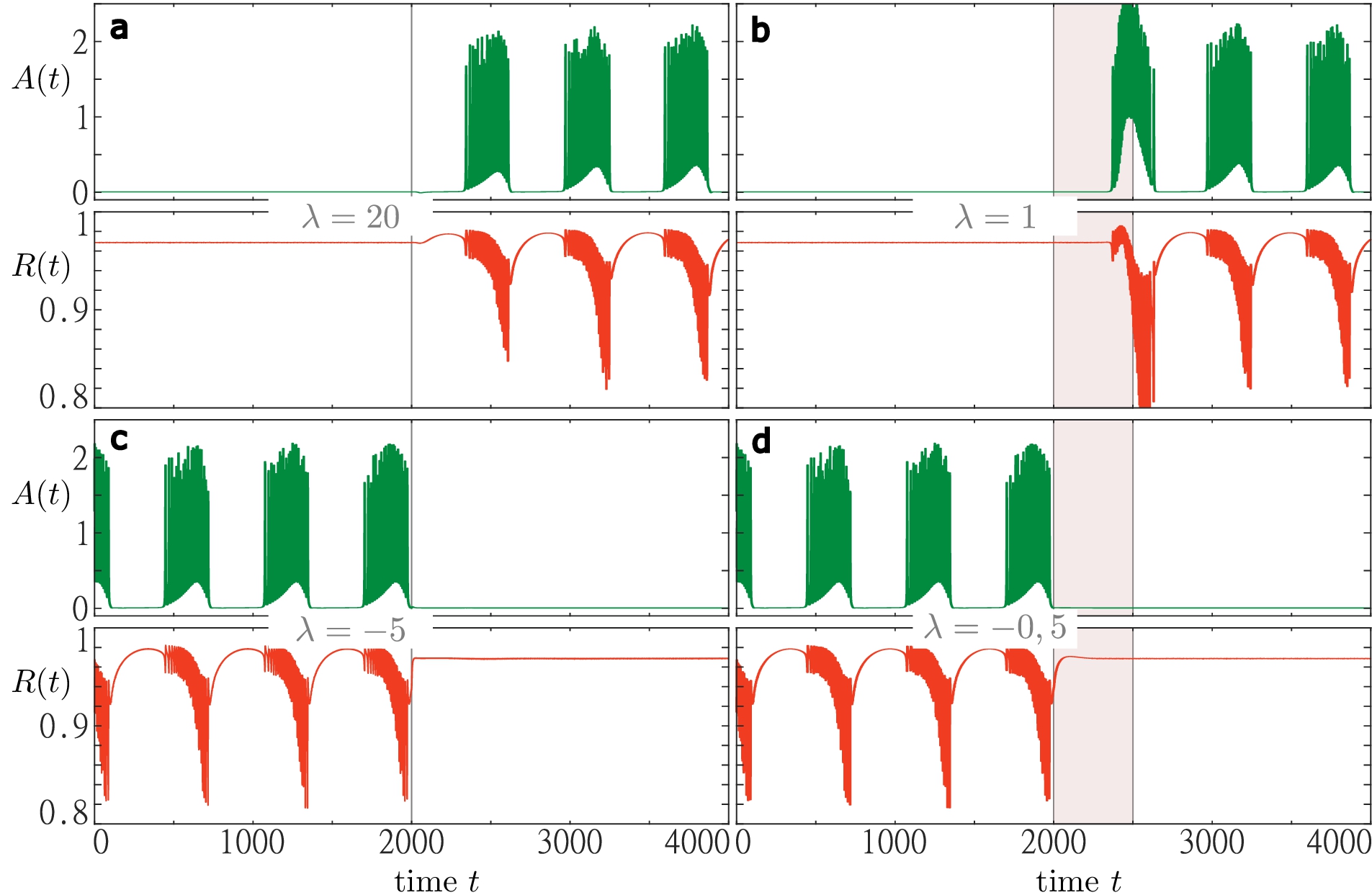}
	\caption{Two perturbation scenarios to induce a switch between an inactive steady state and collective activity bursting. Time traces of macroscopic activity $A(t)$ and order parameter $R(t)$ are shown green and red, respectively. In the panels a and b (c and d), we start from an initial steady state (bursting state). The first perturbation scenario is illustrated for the cases where the resource activity variable $\lambda$ is set to $\lambda=20$ (panel a) and $\lambda=-5$ (panel c) at $t=2000$. The second perturbation scenario is demonstrated for the cases where the resource activity is kept fixed at $\lambda=1$ (panel b) and $\lambda=-0.5$ (panel d) for a duration of $500$ t.u. beginning at $t=2000$ t.u. Simulations were run for $s_1=0.97$ and the remaining parameters fixed as in Figure~\ref{fig:sweeps}.}\label{fig:switching}
\end{figure}
Figure~\ref{fig:switching} shows the results for two different perturbation approaches to
system~\eqref{eq:activeRotModel}--\eqref{eq:lambdaDyn}. The first approach aims to induce a switch in population dynamics by an instantaneous resetting of the resource activity $\lambda$. In the second approach, we induce such a transition by maintaining the resource activity at a certain level for a certain period of time.

The first approach works well for large resetting values of $\lambda$, see Figure~\ref{fig:switching}(a,c). Small values, however, would not be sufficient to induce the macroscopic regime switch. Furthermore, in case of an initial bursting state, eliciting the switch to a steady state depends on the moment at which the perturbation is applied. %\rem{IF}{This is interesting. Is it phase-sensitive excitability of the bursting state?}\rem{RB}{If you call bringing the bursting back to the steady state "excitable" it might be the case :)}.
However, in our numerical simulations (not shown), we have always been able to induce a switching for sufficiently large resetting values of $\lambda$.

Due to the functionally very different nature of the two stable states, there might be reasons to favor one over the other in light of potential applications in medicine. Therefore, it is of great interest to understand simple mechanisms that would induce a switch to the desired state. While the first perturbation approach provides such a mechanism, it still requires strong perturbations which might be undesirable for certain medical reasons, e.g. side effects. Therefore, we have proposed another perturbation approach that leads to a switch while keeping the reset level lower. For this approach, we have also been able to induce switches between a steady state and a bursting state in one or the other direction, see Figure~\ref{fig:switching}(b,d), with the advantage of having the resetting level of the resource activity much lower than for the first method.

In this section, we have proposed two simple perturbation approaches to induce a switch between the two functionally different macroscopic states of the full system which emerge near the transition in layer dynamics between the population activity and inactivity and due to an adaptive dynamical pool of resources. We note that the approaches we proposed are not the only way to induce macroscopic regime shifts. One might also think of perturbing the resource variables ($r_1,r_2$) or even the whole population. Thus, perturbation of the resource activity variable is perhaps the simplest but not the only approach possible. 

%--------------------------------------------------
% Conclusions
%--------------------------------------------------
\section{Conclusions}\label{sec:conclusions}

We have investigated collective dynamics in a system of interacting excitable units coupled to a pool of resources with nontrivial dynamics. The feedback of the resources to the population of coupled excitable units has been realized by an adaptation of the individual units' inputs, whereas in turn, the excitable population is capable of activating or deactivating the pool of resources depending on the population's own activity. As a prototype of excitable local dynamics, we have considered active rotators. Following the ideas outlined by Roberts et al.~\cite{ROB14}, we have assumed the processes at the pool of resources to occur much slower than the local dynamics of excitable units. As a consequence, we have ended up with a system featuring multiscale dynamics, allowing us to use the methods from singular perturbation theory~\cite{DES12,KUE15}.

As our most important finding, we have reported on the phenomenon of collective activity bursting. The phenomenon is characterized by a recurrent switching between episodes of quiescence and episodes of activity bursts in the population of active rotators. To gain a better understanding of the emergence of collective activity bursting, we have made use of the explicit slow-fast timescale separation. In particular, we have divided the system dynamics into the fast layer dynamics of the population and the slow average dynamics of the resources. 

Using the Ott-Antonsen approach, we have analyzed the stability and bifurcations of the stationary solutions of layer dynamics in the thermodynamic limit. For the population of active rotators with a heterogeneity given by a Gaussian distribution, we have derived a bifurcation diagram for the steady state solutions. The bifurcations of layer dynamics depending on the mean and the width of the Gaussian distribution have been corroborated by numerical simulations of a large ensemble of rotators. Doing so, we have determined the parameter regions admitting high or low (or even no) population activity and have obtained the critical lines separating these regions.

Taking the analysis of the layer problem into account, we have further analyzed how the slow averaged dynamics of the resources gives rise to a slow variation of the mean and width of the Gaussian distribution. We have observed the onset of collective activity bursting close to criticality where the population of active rotators undergoes a transition from an inactive to an active state. The emergence of collective bursting is due to a subtle interplay of co-activation and co-deactivation of the dynamical population of rotators and the pool of resources. 

We have further found a region of bistability between collective activity bursting and an inactive steady state close to criticality of the layer dynamics. A similar observation has been also discussed in the context of collective bursting induced by synaptic short-term plasticity~\cite{GAS20}. Moreover, we have proposed two different perturbation methods that can trigger switches between coexisting macroscopic regimes. In particular, we have demonstrated that the regime shifts can be induced either by using instantaneous large perturbations or persistent perturbations of the resource activity. 

In terms of theory, an important extension of our work could concern a further analytical study of the reduced slow-fast system governing the collective dynamics of the ensemble of excitable units and its interaction with the resources. For convenience, we summarize the reduced system here
\begin{align*}
    \dot{z}(I,t)&=\frac{1}{2}(1-z^2(I,t))+iIz(I,t)+\frac{\sigma}{2}Z(t)-\frac{\sigma}{2}\bar{Z}(t)z^2(I,t), \\ 
    \dot{\bm{r}}(t) &= f(\bm{r}(t)-\bm{s},\lambda(t)),\\
    \dot{\lambda}(t) &= -\lambda(t) + \lambda_0 + \rho (r_1 - \text{Im}(Z(t))),
\end{align*}
with
\begin{align*}
    Z(t)=\int g(I)z(I,t)\,\mathrm{d}I.
\end{align*}
%\rem{SE}{Besides changing to $\lambda_0$ i also dropped the time dependence on it}
In a broader context, we have proposed a simple paradigmatic model to study the emergence of complex collective phenomena induced by a dynamically co-evolving pool of resources. The research on the impact of resource constraints on the dynamical regimes of populations of neurons or neuron-like units from the dynamical network perspective~\cite{KRO21,NIC17} has begun only recently. In our study, we have shown that even a simple model that includes nontrivial dynamical resources gives rise to the emergence of collective activity bursting close to criticality in a population of neuron-like excitable units. Our study underlines the potentially important role of resource constraints in the operating of the human brain that is often hypothesized to operate close to criticality. We have further shown that the collective activity bursting may stably coexist with a steady state. Either one of these regimes could be desirable or undesirable, which makes understanding of the control mechanisms to switch between the regimes highly important~\cite{TAN18}. In this context, we have discussed two simple approaches that can successfully induce such regime shifts. Both approaches impose perturbations to the single activity variable of the resource pool and can thus be generalized to systems with even more complex dynamical resource pools.

\section*{Conflict of Interest Statement}
%All financial, commercial or other relationships that might be perceived by the academic community as representing a potential conflict of interest must be disclosed. If no such relationship exists, authors will be asked to confirm the following statement:
The authors declare that the research was conducted in the absence of any commercial or financial relationships that could be construed as a potential conflict of interest.

%\section*{Author Contributions}
%All authors listed have made a substantial, direct, and intellectual contribution to the work and approved it for submission.

\section*{Funding}
The work of R.B. and S.Y. was supported by the German Research Foundation DFG, Project Nos. 411803875 and 440145547. I.F. acknowledges funding from the Institute of Physics Belgrade through the grant by the Ministry of Education, Science and Technological Development of the Republic of Serbia. We acknowledge support by the German Research Foundation (DFG) and the Open Access Publication Fund of Humboldt-Universität zu Berlin.

%\section*{Acknowledgments}
%Anyone?

%\section*{Supplemental Data}
 %\href{http://home.frontiersin.org/about/author-guidelines#SupplementaryMaterial}{Supplementary Material} should be uploaded separately on submission, if there are Supplementary Figures, please include the caption in the same file as the figure. LaTeX Supplementary Material templates can be found in the Frontiers LaTeX folder.

%\section*{Data Availability Statement}
%The datasets [GENERATED/ANALYZED] for this study can be found in the [NAME OF REPOSITORY] [LINK].
% Please see the availability of data guidelines for more information, at https://www.frontiersin.org/about/author-guidelines#AvailabilityofData

%\bibliographystyle{prwithtitle}
%\bibliography{FRA22_refs}
%\bibliography{references-sy}
%\bibliography{refs,references-sy,ref_if}

\end{document}